\documentclass[twocolumn,prl,showpacs,floatfix]{revtex4}
\usepackage{graphicx}
\usepackage{epsfig}
\usepackage{rotating}
\usepackage{amsmath}
\usepackage{amsfonts}
\usepackage{amssymb}
\usepackage{enumerate}
\usepackage{longtable}
\usepackage{dcolumn}% Align table columns on decimal point
\usepackage{bm}

\begin{document}

\title{Neutron spectra of Herbertsmithite Materials: Observation of a Valence Bond Liquid phase?}

\author{R. R. P. Singh}
\affiliation{University of California Davis, CA 95616, USA}

\date{\rm\today}

\begin{abstract}
We argue that neutron spectra in a short-range ordered Valence Bond
state is dominated by two-spinon states localized in small spatial
regions such as a pinwheel. These excitations lead to angle averaged
dynamic structure factor that is spread over a wide frequency range
up to about $2.5 J$, whereas its wavevector dependence at all
frequencies remains very close to that of isolated dimers. These
results are in excellent agreement with recent Neutron scattering data
in the Herbertsmithite materials ZnCu$_3$(OH)$_6$Cl$_2$.
\end{abstract}

%\pacs{}

\maketitle
Recent experimental studies\cite{shores,helton,shlee,olariu,imai,devries1} of the Herbertsmithite material ZnCu$_3$(OH)$_6$Cl$_2$ 
with structurally perfect Kagome planes has 
brought renewed interest in the study of Quantum Spin-Liquid phases in the Kagome Lattice 
Heisenberg Model.\cite{pwa,elser,mila,misguich,palee,mpaf} The issues of Quantum
Spin-Liquid versus Valence Bond
Crystal Order,\cite{VBC} of a possible gap in the spin excitation spectra and of 
deconfinement of fractional spin excitations 
continue to be subjects of intense theoretical
debate.\cite{ed,dmrg,singh-huse,ybkim,ms07,laeuchli,vidal,tchernyshyov,poilblanc} The
experimental studies provide a complimentary perspective to this long standing problem.

In a recent letter de Vries et al presented neutron scattering data on these 
materials over a range of momentum and frequency transfers.\cite{devries} 
They discuss their data primarily in the context of Algebraic Spin-Liquid and other 
theories.\cite{palee,mpaf}
Here, we would like to argue that the data is much better understood in terms of a 
Valence Bond phase\cite{VBC,singh-huse} with well developed short-ranged Valence Bond order but no
long-range Valence Bond Crystal order. Such a finite temperature
phase, lacking long-range quantum coherence,
may appropriately be called a classical Valence Bond Liquid.

The powder-diffraction neutron data of de Vries et al covers a temperature range from 2K to 120 K,
an energy transfer of up to $30$ meV and the full range of momentum transfer values.
Their key findings can be summarized as follows: After allowance is made for
phonons as well as for some impurity spins, the magnetic behavior intrinsic to the
system, is rather well described by an angle averaged q-dependent form-factor
which is essentially identical to that of a single spin-dimer. However, unlike a single
dimer, where such a spectrum would be strongly peaked at the singlet-triplet energy
gap ($J$ for an isolated dimer), the spectral weight is spread nearly uniformly
over a wide range of energies extending at least down to $4$meV and up beyond $30$ meV.
Furthermore, this behavior is evident at 2K but persists also at 120 K.
The frequency dependence makes the behavior clearly inconsistent with that of isolated
dimers and the temperature dependence makes it inconsistent with a Valence Bond Crystal
order (and any other quantum ground state with an ordering scale much smaller than 100K
as expected in the model).

\begin{figure}
\begin{center}
\includegraphics[width=0.9\columnwidth,clip,angle=0]{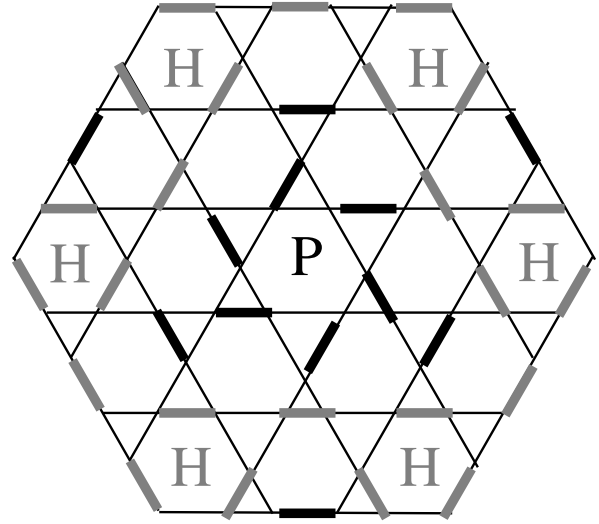}
\caption{\label{fig:Fig1} Proposed Valence Bond Phase of the Kagome Lattice
Heisenberg Model consists of a Honeycomb Lattice of resonating hexagons (H),
where six hexagons surround a pinwheel (P). The pinwheels are isolated from empty triangles
leading to substantially reduced quantum fluctuations on their dimer bonds.
Triplets on dimers represented by black thick lines are heavy nearly immobile particles, 
whereas triplets on grey thick lines represent particles mobile throughout the lattice.
}
\end{center}
\end{figure}

We first note that the exchange constant for the material has
been estimated\cite{helton,rigol07,sindzingre07} to be
in the range 170K-190K, which translates to about $15$meV. 
There is increasing theoretical support for a Valence Bond Crystal (VBC) ground state
of the Kagome-Lattice Heisenberg Model with a 36-site
unit cell.\cite{singh-huse,vidal} One of the distinctive features
of the VBC state is a pin-wheel (See Fig.~1) in each unit cell, 
a defect free structure of Valence-Bond
containing triangles, where the bonds remain almost completely dimerized. 
An isolated pinwheel is an example of a 
Delta chain,\cite{kubo,shastry,tchernyshyov}
a system of corner sharing triangles, where the Hamiltonian is minimized by 
dimerization.\cite{majumdar} In the VBC state, their is significant quantum
fluctuations. In particular, the dimerization inside the resonating hexagons is
strongly reduced. But, the pinwheel region is geometrically protected against
quantum fluctuations and hence remains essentially fully dimerized.\cite{singh-huse2}

Different Valence Bond phases have energy difference of only $0.001$ J per site.\cite{singh-huse}
Thus one expects any phase transition to such a state to occur at a very low temperature.
However, short-range Valence Bond order is set by $J$ and can develop at
significantly higher temperatures. Once this short-range order is a few lattice
constants, pin-wheel like structures should begin to form locally. Because they
have extremely low local energies, they should be stable up to higher temperatures.

\begin{figure}
\begin{center}
\includegraphics[width=0.9\columnwidth,clip,angle=0]{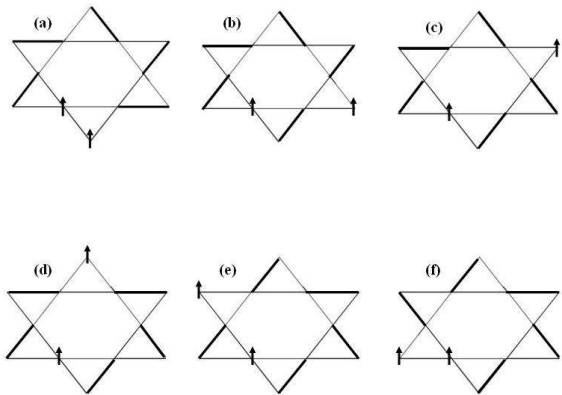}
\caption{\label{fig:Fig2} Kink-antikink or 2-spinon states for the pinwheel.
A triplet excitation is created by breaking a singlet bond in the ground state.
The Kink spinon lies on the inner hexagon, while the more mobile antikink
spinon can move on the outside vertices of the pinwheel through states (a) through
(f) shown in the figure. When the two spinons are together as in Fig.~(a),
the kink spin can also move by going through higher
energy intermediate states.}
\end{center}
\end{figure}

The triplet excitations of
the pinwheel are kink-antikink pairs.\cite{kubo,shastry,tchernyshyov}  
In an infinite Delta chain, the kinks are immobile, where as the antikink
can hop from one triangle to another. The kink has zero excitation energy,
whereas the energy of the antikink can be approximated by
\begin{equation}
\epsilon(k)=5/4-\cos{k}.
\end{equation}
Note that this means that excitations extend over the energy range 
$J/4<\epsilon<9 J/4$. For the infinite system more detailed analytical
and numerical calculations show\cite{kubo,shastry} that the lowest energy antikink is
roughly at $0.219 J$, whereas the upper energy may extend up to as
much as $3J$. 
For the finite system, spin excitation from the ground state
creates a pair of parallel spins on one of the singlet bonds of the ground
state. While the kink antikink description is roughly valid, because
both the kink and antikink remain in each other's vicinity, both
free spins or spinons become mobile. 
Fig.~2, shows the configurations that correspond to the mobile antikink.
On the other hand the kink can also move by going through higher energy
intermediate configuration, when the two spins are next to each other.
In the $S^z=1$ sector, assuming two free spins and the rest of the system in
the ground state leads to 66 states, of which
$36$ states are of the kink-antikink type, where
one spin is in the inner hexagon, whereas the other spin is
on the outside vertices of the pinwheel. One expects the spectral
weight to be primarily spread over these states, giving rise
to spectral weight spread roughly over the energy range $J/4<\epsilon< 9J/4$. 

\begin{figure}
\begin{center}
\includegraphics[width=0.8\columnwidth,clip,angle=270]{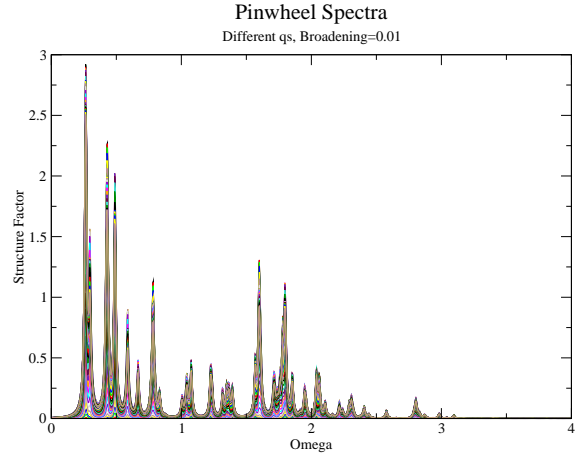}
\caption{\label{fig:Fig3} Angle averaged dynamic structure factor versus
frequency of the Pinwheel
state at $T=0$ with small broadening.}
\end{center}
\end{figure}

\begin{figure}
\begin{center}
\includegraphics[width=0.8\columnwidth,clip,angle=270]{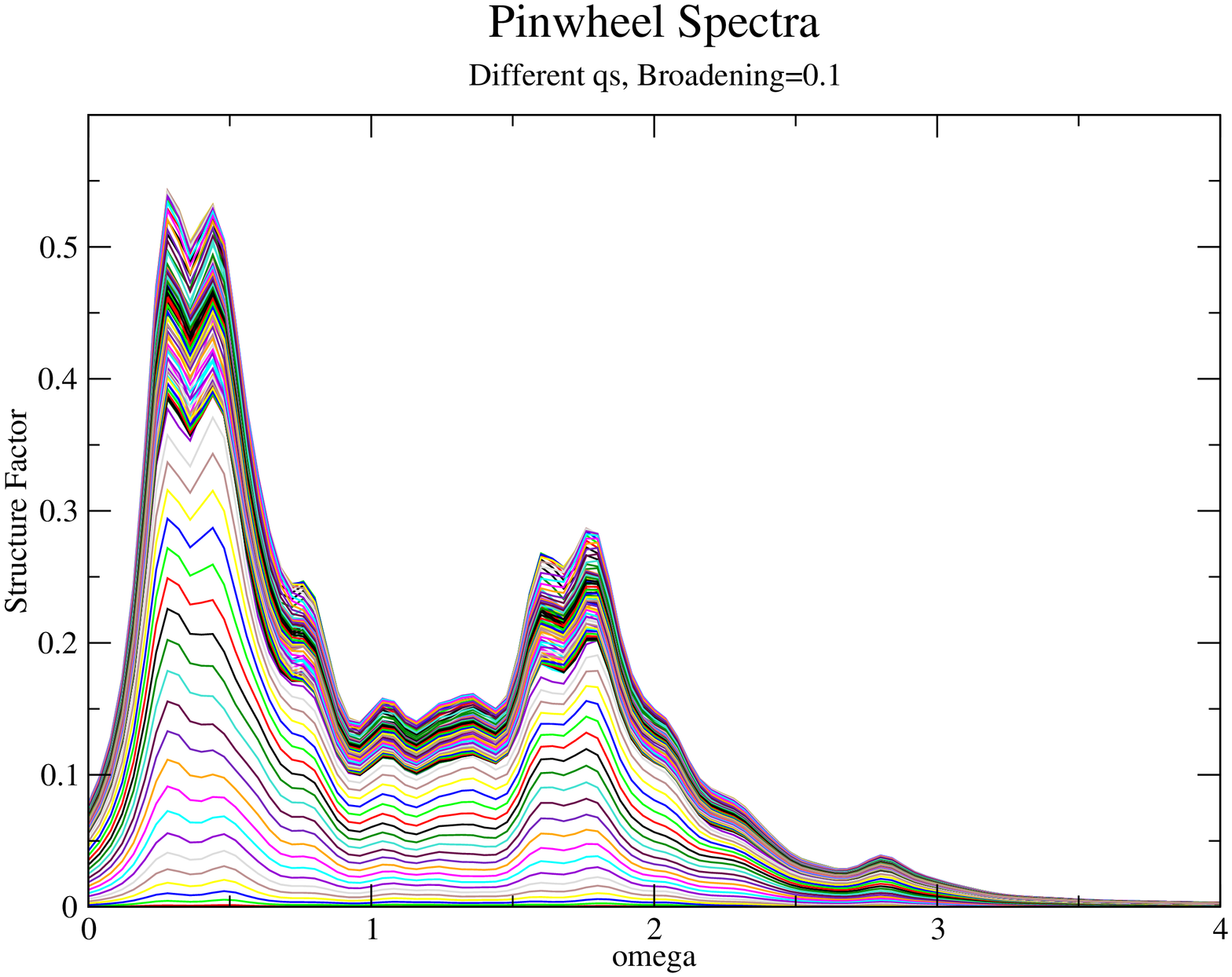}
\caption{\label{fig:Fig4} Angle averaged dynamic structure factor 
versus frequency of the Pinwheel
state at $T=0$ with larger broadening.}
\end{center}
\end{figure}

\begin{figure}
\begin{center}
\includegraphics[width=0.8\columnwidth,clip,angle=270]{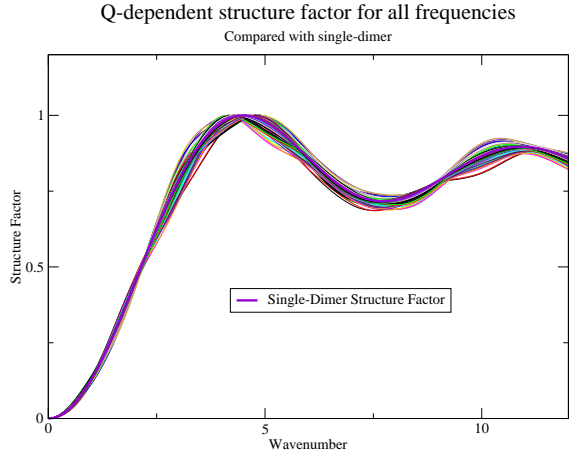}
\caption{\label{fig:Fig5} Angle averaged dynamic structure factor 
at different frequencies, scaled to have a maximum of unity, versus wavenumber
compared with results of a single dimer.}
\end{center}
\end{figure}

This can be easily confirmed by exact diagonalization of the 12-site Heisenberg model
on a pinwheel. One finds that the
lowest triplet state has an excitation energy of $0.260 J$. 
The spectral weight is spread over a large number of states.
The highest spectral weight of any one single state is only about $5$ percent.
The states with the highest 36 spectral weights are spread over the energy range
$0.260 J<\epsilon<2.039J$ and they contribute above 80 percent to the spectral
weight.
%The highest contributing $66$ states are spread over an energy range
%of $0.260 J<\epsilon<2.809 J$ and these states have 96 percent of the total 
%spectral weight. 
If we look at total spectral weight up to some energy, 
roughly $95.5$ percent of the weight extends up to an energy of $2.25$ J, 
$97.7$ percent of the weight extends up to an energy of $2.5 $ J and roughly
$99.7$ percent of the weight extends up to an energy of $3$ J.
This strongly confirms that the dominant spectral contributions
come from the kink-antikink states.

In the Valence Bond Crystal state, these triplets can ultimately
decay to still lower lying light triplets,\cite{singh-huse2} though those decay times
are likely to be very long, because the pinwheels are surrounded by dimerized triangles
and have no empty triangles in their immediate vicinity, which strongly reduces quantum
fluctuations.
In the liquid phase, these excitations should have a shorter finite 
lifetime, which one could represent by a Lorentzian broadening. 
The frequency dependence
of the angle averaged dynamic structure factor, for several q-values,
at a small Lorentzian broadening  (0.01 J) is shown in Fig 3, where as at a 
larger broadening
(0.10 J) it is shown in Fig. 4. The latter may be more representative of
the liquid state. One finds that at higher broadening
one has a spectral weight that is
spread roughly continuously up to an energy of about  $3J$. 

For any pair of spins at a distance $r$, the angle averaged value of
$\exp{(i\vec q\cdot\vec r)}$ is given by $\sin{(qr)}/qr$. This can be used
to calculate the angle averaged dynamic structure factor for any given $q$.
For all frequencies,
the structure factor as a function of wavenumber shows dependence, which
is very close to that of an isolated dimer (See Fig. 5). In an isolated
pinwheel, the equal-time correlation function is strictly that of a dimer,
but excitations are extended over the full pin-wheel. Thus the energy
integrated structure factor is strictly that of an isolated dimer, but not
the spectral weight at a given energy. However, what the calculations show
is that while the spectral weight is spread out over a wide range
of frequencies, the q-dependence is always near that of an isolated dimer.
Since delta chains are likely to be ubiquitous in the short-range 
Valence Bond ordered phase of the Kagome Lattice Heisenberg model,\cite{tchernyshyov} this spectral feature may persist up to energy comparable to $J$,
that is, as long as the system has short range Valence Bond order.

This picture implies that the equal-time spin-spin correlations in the system are
essentially only nearest neighbor. However, the arrangement of dimerized triangles
makes any triplet excitation break into a kink-antikink pairs. These lead to
spectral weight spread over a wide frequency range.

In fact, the spectra obtained by exact-diagonalization of 24, 30 and 36-site
clusters by Laeuchli and Lhuillier\cite{laeuchli} show very similar frequency dependence, where much
of the spectral weight is nearly uniformly spread between $J/4$ to about $2.5 J$.
The finite size system has only short-range Valence bond order.
This is further evidence that almost all states with short range Valence Bond order have
these features.

One also knows that the Herbertsmithite materials have a small Dzyaloshinski-Moria
anisotropy.\cite{rigol,mendels,cepas} These anisotropies would strongly
influence the low energy spectra and the nature of long-range order
without significantly altering the high energy spectral properties.
In the experiments, the reduction in spectral weight at low temperatures
below an energy of $4$ meV may well be related to the DM anisotropy which
would cause the low energy spectral weight to move to even lower energies.

Furthermore, the true signature of long-range VBC order, would be
the observation of low energy light triplets, which live on the perfect
hexagon and bridging dimers.\cite{singh-huse2,ybkim} In the VBC phase
these provide an extended network for the triplets to move around,
whereas the pinwheels form isolated pristine regions which are
nearly protected from quantum fluctuations, and have only localized
triplets. The honeycomb VBC state
has 50 percent of the spectral weight in the non-fluctuating
dimer-like heavy excitations and the rest of the 50 percent of the spectral
weight in the light mobile triplets, which with long-range VBC order
have an energy gap of order $J/20$.\cite{ed,dmrg}

In conclusion, we have argued that the observation of dimer-like q-dependence combined with
a spectral weight spread over a wide frequency range and remaining nearly
temperature independent over a wide range of temperature is strongly suggestive
of a Valence Bond Liquid phase, with short-range Valence Bond Order exceeding
a couple of lattice constants. These excitations can be regarded as a
kink-antikink pair or two spinons, which are confined in a very small
spatial region. They extend up to fairly high temperatures of order $J$,
where such excitations are necessarily broadened by the underlying
classical liquid environment. Whether at low enough temperatures, the
development of long-range quantum coherence leads to truly delocalized
spinons as in algebraic quantum spin-liquid phases
or to the development of much sharper localized and extended triplet excitations as in
a Valence Bond Crystal remains to be seen.

\begin{acknowledgements}
We would like to thank Mark de Vries
for valuable discussions.
\end{acknowledgements}

%\bibliographystyle{apsrev}
%\bibliography{../bibinput/liter10}

\end{document}